\documentclass[a4paper,11pt]{article}%

\usepackage{amsfonts}
\usepackage{graphicx}
\usepackage{amsmath}
\usepackage{amssymb}%
\usepackage{cite}
\pdfoutput=1
\makeatletter

\@addtoreset{equation}{section}
\makeatother
\newcommand{\beq}{\begin{equation}}
\newcommand{\eeq}{\end{equation}}
\renewcommand{\a}{\alpha}
\renewcommand{\b}{{{\beta}}}

\newcommand{\be}{\begin{eqnarray}}
\newcommand{\ee}{\end{eqnarray}}

\usepackage[utf8]{inputenc}
\usepackage{amsmath}
\usepackage{amscd}
\usepackage{amsfonts}
\usepackage{amssymb}
\usepackage[english]{babel}
\usepackage{braket}
\usepackage{mathtools}
\usepackage{amsthm}
\usepackage{graphicx}
\usepackage{graphicx,color}
\usepackage{enumerate}
\usepackage{verbatim}
\usepackage[all]{xy}
\usepackage{xcolor}
\usepackage{comment}

\newcommand{\virgolette}{``}

\numberwithin{equation}{section}

\theoremstyle{plain}

\theoremstyle{definition}

\theoremstyle{remark}

\begin{document}
\baselineskip=18pt
\baselineskip 0.6cm

\begin{titlepage}
\setcounter{page}{0}
\renewcommand{\thefootnote}{\fnsymbol{footnote}}
\begin{flushright}
ARC-19-25 
\\
\today
\end{flushright}

\vskip 2cm
\begin{center}
{\Huge \bf  Super Chern-Simons Theory:} \\\vskip .2cm
{\Huge \bf  BV-formalism and $A_\infty$-algebras}
\vskip 1cm
{
\large {\bf C.~A.~Cremonini}$^{~a,b}$\footnote{carlo.alberto.cremonini@gmail.com} 
and
\large {\bf P.~A.~Grassi}$^{~c,d,e}$\footnote{pietro.grassi@uniupo.it}
}
\vskip .5cm {
\small
\medskip
\centerline{$^{(a)}$ \it Dipartimento di Scienze e Alta Tecnologia (DiSAT),}
\centerline{\it Universit\`a degli Studi dell'Insubria, via Valleggio 11, 22100 Como, Italy}
\medskip
\centerline{$^{(b)}$ \it INFN, Sezione di Milano, via G.~Celoria 16, 20133 Milano, Italy} 
\medskip
\centerline{$^{(c)}$
\it Dipartimento di Scienze e Innovazione Tecnologica (DiSIT),} \centerline{\it Universit\`a del Piemonte Orientale, viale T.~Michel, 11, 15121 Alessandria, Italy}
\medskip
\centerline{$^{(d)}$
\it INFN, Sezione di Torino, via P.~Giuria 1, 10125 Torino, Italy}
\medskip
\centerline{$^{(e)}$
\it Arnold-Regge Center, via P.~Giuria 1,  10125 Torino, Italy}
\medskip
}
\end{center}
\vskip 0.2cm
\centerline{{\bf Abstract}}
\medskip
	This is a companion paper of a long work appeared in \cite{Cremonini:2019aao} discussing 
the super-Chern-Simons theory on supermanifolds. Here, it is emphasized that the BV formalism is naturally formulated 
using integral forms for any supersymmetric and supergravity models and we show how to deal with 
$A_\infty$-algebras emerging from supermanifold structures. 


\end{titlepage}

\tableofcontents \noindent {}

\newpage
\setcounter{footnote}{0} \newpage\setcounter{footnote}{0}


\section{Introduction}

The BV-formalism and supergeometry have been extensively studied during the last years. It has been shown the naturalness 
of the BV-formalism in the supergeometry approach (QP-manifolds, odd-symplectic structures, BV bracket) because all fields 
have their own (classical) opposite-statistic partner leading to a BV-symplectic 2-form corresponding to an odd-symplectic 
structure (see \cite{Khudaverdian:1995np,Khudaverdian:2000zt,Khudaverdian:2006agq} and 
\cite{Alexandrov:1995kv}). 
The application of the BV-formalism was ubiquitous in quantum field theory and string theory, but in our opinion 
the BV-formalism for supersymmetric theories has never been deeply explored from the supergeometric point of view and 
this is the aim of the present work. 

As has been shown some years ago by several authors \cite{Castellani:2015ata,CCGGeom,CCGir,Grassi:2016apf}, 
any supersymmetric model can be reformulated on 
a given supermanifold by constructing a $p$-form Lagrangian ${\cal L}^{(p)}$ (the rheonomic Lagrangian, 
defined according to the rules given in \cite{castellanidauriafre}). The action functional is obtained by multiplying the 
Lagrangian by a PCO $\mathbb{Y}$ (also known as Poincar\'e dual of the immersion of the bosonic submanifold into the supermanifold) 
which converts the  $p$-form Lagrangian ${\cal L}^{(p)}$ into an integral form ${\cal L}^{(p)}\wedge \mathbb{Y}$ which can be 
integrated on the supermanifold. By choosing $\mathbb{Y}$ one can obtain any superspace representation of the same 
supersymmetric action. 

In \cite{Grassi:2016apf}, that procedure has been applied to super-Chern-Simons theory $D=3, N=1$ 
and the details of the construction have been discussed. It has been pointed out that there might be another way 
to describe Chern-Simons theory using pseudoforms (by pseudoforms we intend those forms with a non-zero
number of delta functions less than the maximal one or, differently stated in our case, with picture number equal to one). In that 
case there are some caveats. Indeed, in order to write the interactions, one needs to insert some PCO's (lowering 
the number of picture) which are potential sources of ambiguities and difficulties. 
In \cite{Cremonini:2019aao}, it is shown how to deal with those issues by introducing a suitable 
set of multiproducts leading to an $A_\infty$ algebra.

What is left to study is the BV-formalism in the framework of integral forms. For that reason, we again  
use Chern-Simons theory to pave the ground for more complicated models\cite{Witten:1992fb}. 
We show that the natural way to introduce the antifields in the game is by using the supermanifold 
version of Serre's duality. Then, when working with the theory at zero picture number, the natural set of antifields lies into the integral forms complex instead 
of the usual superforms complex. That automatically takes into account the correct number of degrees of freedom 
needed to use the BV-formalism. In the previous works \cite{Grassi:2016apf,Cremonini:2019aao},  the 
closure of the Lagrangian form is imposed by consistency with the entire construction 
in order to have the full off-shell supersymmetry and to allow any change of the PCO  interpolating among 
any superspace realisations. We show that the closure of the antifield part of the Lagrangian form is easily achieved 
and this is consistent with the antifield formalism. We also show that the antifield part of the Lagrangian can be 
written in a form that is similar to the gauge field part of the action with the PCO's. In particular, this allows us to check that, after specifying suitable choices of PCO's, we get the known superspace or component formulations. 

The last part of the present work concerns the discussion of the BV formalism in the 
language of pseudoforms. In that case, the antifields are introduced as pseudoforms and we get that again the 
multiproducts are needed also in the antifield sector generating the $A_\infty$ structure. We show that 
the BV formalism easily adapts to the present framework and the result leads to CS-action 
where the picture number one gauge field $A^{(1|1)}$ is replaced by a generic pseudoform ${\cal A}$ 
with any form degree (in particular, it can be negative, exactly as in string field theory \cite{Zwiebach:1992ie}) and picture number fixed to one. The action is automatically 
invariant under superdiffeomorphisms and it has the standard gauge symmetry. The couplings are 
obtained by multiproducts, where we have inserted the generic pseudoform ${\cal A}$. 
We show that the action can not be redefined into a trivial one (i.e. without multiproducts) and we show how to 
retrieve the supersymmetry in the present context.


\section{Superforms, Integral forms and Pseudoforms}

The space of differential forms has to be extended in order to define a meaningful integration theory. Given a supermanifold $\mathcal{SM}^{(m|n)}$ of super-dimension  $(m|n)$, we define $\displaystyle \Omega^{(\bullet | \bullet)} \left( \mathcal{SM} \right) $ as the complete complex of forms; they are graded w.r.t. two gradings as $\displaystyle \Omega^{(\bullet | \bullet)} = \oplus_{p,q} \Omega^{(p|q)}$, where $ q = 0 , \ldots , m $, $ p \leq n $ if $ q = m $ , $p \geq 0$ if $q=0$ and $p \in \mathbb{Z}$ if $q \neq 0,m$. The wedge product for form multiplication is used in the paper, with suitable adjustments due to the picture number as shown in what follows.

Locally, a $(p|q)$-form $\omega$ formally reads 
\begin{equation}\label{inNBA}
\omega = \sum_{l,h,q} \omega_{[a_1 \dots a_l] (\a_{1} \dots \a_{h}) [\b_{1} \dots \b_{q}]}  dx^{a_1} \dots dx^{a_l} 
d\theta^{\a_1} \dots d\theta^{\a_h} \delta^{^{g(\beta_1)}}(d\theta^{\b_1}) \wedge   \dots \wedge \delta^{^{g(\beta_q)}}(d\theta^{\b_q}) \ ,
\end{equation}
where $g(x)$ denotes the differentiation degree of the Dirac delta function corresponding to the 1-form $d\theta^x$. The \emph{picture number} $q$ counts the number of Dirac delta functions, while $p$ counts the form number. The 
three indices $l, h$ and $q$ satisfy the relation 
\begin{equation}\label{inNBB}
l + h - \sum_{k=1}^q g(\b_k) = p\,, ~~~~~~\a_l \neq \{\b_1, \dots, \b_q\} ~~~ \forall l=1,\dots,h\, .
\end{equation}
Each $\alpha_l$ in the above summation should be different from any $\beta_k$, otherwise the 
degree of the differentiation of the Dirac delta function could be reduced and the corresponding 1-form $d\theta^{\a_k}$ removed from the basis. 
 The components  $\omega_{[i_1 \dots i_l] (\a_{1} \dots \a_{m}) [\b_{1} \dots \b_{r}]}$  
of $\omega$ are superfields.

Once the integral forms are defined, we have to clarify how the integration is performed: given a \emph{top form} $\omega^{(m|n)}$, i.e. a form with either maximum picture number or maximum form number, we write
\begin{equation}
	I \left( \omega^{(m|n)} \right) = \int_{\mathcal{SM}^{(m|n)}} \omega^{(m|n)} \ .
\end{equation}
The integral is performed by first integrating over $dx$'s, which amounts to selecting the top form, then we use the Berezin integral over $\theta$'s, and the integration over $d\theta$, viewed as 
algebraic bosonic variables \cite{Witten:2012bg,CCGGeom}, is performed as a formal algebraic integration using the distributional properties of $\delta(d\theta)$'s. The final expression 
needs a usual Riemann/Lebesgue integral on $x$'s.

The elements of $\Omega^{(p|0)}$ are denoted by \emph{superforms} and are represented as polynomials of $dx$'s and $d \theta$'s; the forms of the spaces $\Omega^{(p|n)}$ are denoted by \emph{integral forms} and are represented as polynomials of $dx$'s and the product $\delta \left( d \theta^1 \right) \ldots \delta \left( d \theta^n \right)$ and derivatives of the Dirac delta's; finally $\Omega^{(p|q)}$, $0 < q < n$, are denoted as \emph{pseudoforms}.\footnote{Notice that the definition of pseudoforms in Voronov et al. is slightly different \cite{VorGeom,Voronov:1999mrk}.} At a given form number, $\Omega^{(p|0)}$  and $\Omega^{(p|n)}$  are finite-dimensional spaces, while $\Omega^{(p|q)}$ are infinite dimensional spaces.

Besides the wedge product, we recall that the spaces of forms $\Omega^{(p|q)}$ admit a differential $d$ acting as an antiderivation on each single space and analogously we can define the contraction operator $\iota_X$, where $X$ is a vector field. Notice that if $X$ is an odd vector field, $\iota_X$ is a commuting derivation. There are several new operators, ($Z , \mathbb{Y} , \Theta , \eta$) acting on forms, inspired by string theory, which modify also the picture number. In particular let us consider $\mathbb{Y}$: given a $(p|q)$-form $\displaystyle \omega^{(p|q)} \in \Omega^{(p|q)}$, we define the \emph{Picture Raising Operator} $\mathbb{Y}^{(0|s)}$ as a multiplicative operator s.t.
	\begin{eqnarray}
		\label{PCOY} \mathbb{Y}^{(0|s)} : \Omega^{(p|q)} \longrightarrow \Omega^{(p|q+s)} \ , \ \omega^{(p|q)} \mapsto \omega^{(p|q)} \wedge \mathbb{Y}^{(0|s)} \ ,
	\end{eqnarray}
	i.e. it raises the picture number by $s$.

Again, given a $(p|q)$-form $\displaystyle \omega^{(p|q)} \in \Omega^{(p|q)}$, we define the \emph{Picture Lowering Operator} $Z_D$ as
	\begin{eqnarray}
		\label{PCOZ} \nonumber Z_v : \Omega^{(p|q)} \longrightarrow \Omega^{(p|q-1)} \ , \ \omega^{(p|q)} \mapsto Z_v \left( \omega^{(p|q)} \right) = \left[ d , -i \Theta ( \iota_D ) \right] \omega^{(p|q)} \ ,
	\end{eqnarray}
	where $[ \cdot , \cdot ]$ denotes as usual a graded commutator and the action of the operator $\Theta ( \iota_v )$ is defined by the Fourier-like relation of the Heaviside step function
	\begin{equation} \label{PCOT}
		\Theta ( \iota_v ) \omega^{(p|q)} ( d \theta^\alpha ) = - i \lim_{\epsilon \to 0} \int_{- \infty} ^\infty \frac{dt}{t + i \epsilon} e^{i t \iota_v} \omega^{(p|q)} \left( d \theta^\alpha \right) = - i \lim_{\epsilon \to 0} \int_{- \infty} ^\infty \frac{dt}{t + i \epsilon} \omega^{(p|q)} \left( d \theta^\alpha + i t v^\alpha \right) \ ,
	\end{equation}
	where we have used the fact that $\displaystyle e^{i t \iota_v} $ is a translation operator. Hence the operator $\Theta (\iota_v)$ maps
	\begin{equation*}
		\Omega^{p|q} \to \Omega^{p-1|q-1} \ ,
	\end{equation*}
	i.e. it lowers either the form degree or the picture degree. As has been shown in \cite{Cremonini:2019aao} this operator does not give a pseudoform as a result, but rather an $inverse$ $form$, i.e. an expression containing negative powers of $d \theta$ \cite{Catenacci:2018xsv}. The computation techniques and results are contained in \cite{Cremonini:2019aao}. There,
	it has been described the $\eta ( \iota_v )$ operator as well. The latter is crucial to defined and build the higher products of the 
	$A_\infty$ algebra.

\section{Super Chern-Simons Actions (SCS)}

\subsection{At Picture Zero} 

We briefly recall some basic facts about $D=3$, ${\cal N}=1$ Super Chern-Simons theory. That model serves as 
a simple playground for more sophisticated examples.  
We start from a $(1|0)$-superform $A^{(1|0)} = A_a V^a  + A_\a \psi^\a$, (where the superfields $A_a(x,\theta)$ and $A_\alpha(x,\theta)$ take values in the adjoint representation of the gauge group) and we define the field strength
\begin{eqnarray}\label{SM-A}
F^{(2|0)} = d A^{(1|0)} +  A^{(1|0)} \wedge A^{(1|0)} = F_{[ab]} V^a \wedge V^b + F_{a\alpha} V^a \wedge \psi^\a + F_{(\a\b)} \psi^\a\wedge \psi^\b \,.
\end{eqnarray}
In order to reduce the redundancy of degrees of freedom of $A_a$ and $A_\a$ of the $(1|0)$-form $A^{(1|0)}$, one imposes a priori the \emph{conventional constraint}
\begin{eqnarray}
\label{SM-C}
\iota_\a \iota_\b F^{(2|0)} =0\,   ~~~ \Longleftrightarrow ~~~ F_{(\a\b)} = D_{(\a} A_{\b)} + \gamma^a_{\a\b} A_a +\{A_\alpha, A_\beta\} =0\,, 
\end{eqnarray}
from which it follows that $F_{a\alpha} = \gamma_{a, \alpha\beta} W^\beta$ with 
$W^\a = \nabla^\b \nabla^\a A_\b$ and $\nabla_\a W^\a =0$. 
The gaugino field strength $W^\a$ (a $(0|0)$-form) is gauge invariant 
under the non-abelian transformations $\delta A_\a = \nabla_\a \Lambda$. 
These gauge transformations descend from the gauge transformations of $A^{(1|0)}$, $\delta A^{(1|0)} = \nabla \Lambda$ where $\Lambda$ is 
a $(0|0)$-form.  

In order to express the action as an integral on a supermanifold, we use the Poincar\'e dual form (as known as PCO) $\mathbb{Y}^{(0|2)}$ dual to the immersion 
of ${\cal M}^{(3)}$ into ${\cal SM}^{(3|2)}$.   The Poincar\'e dual form $\mathbb{Y}^{(0|2)}$ is closed, 
it is not exact and any of its variation is $d$-exact. 
The action can now be written on the full supermanifold as
\begin{eqnarray}
\label{reC}
S[A] = \int_{{\cal SM}^{(3|2)} } {\cal L}^{(3|0)}(A, dA)\wedge  \mathbb{Y}^{(0|2)}\,.
\end{eqnarray}

Any variation of the embedding yields $\delta \mathbb{Y}^{(0|2)} = d \Lambda^{(-1|2)}$ and leaves the action invariant if the Lagragian is closed. The rheonomic Lagrangian ${\cal L}^{(3|0)}(A, dA)$ reads
\begin{eqnarray}
\label{reD}
{\cal L}^{(3|0)}(A, dA) = {\rm Tr}\Big( A^{(1|0)}\wedge dA^{(1|0)} + \frac23 A^{(1|0)}\wedge A^{(1|0)}\wedge A^{(1|0)} + W^{(0|0)\a} \epsilon_{\a\b} W^{(0|0)\b} V^3  \Big)\wedge {\mathbb Y}^{(0|2)}\,, 
\end{eqnarray}
which is a $(3|2)$ form, $V^3 = \frac{1}{3!}\epsilon_{abc} V^a \wedge V^b\wedge V^c$.

This is the most general action and the closure of ${\cal L}^{(3|0)}$ implies that any gauge invariant 
and supersymmetric action can be built by choosing a PCO $\mathbb{Y}^{(0|2)}$ inside the same cohomology class. Therefore, 
starting from the rheonomic action, one can choose a different ``gauge" -- or better said a different immersion 
of the submanifold ${\cal M}^{(3)}$ inside the supermanifold ${\cal SM}^{(3|2)}$ -- leading to different form of the action with the same physical content. 


\subsection{At Picture One}

Any PCO $\mathbb{Y}^{(0|2)}$ can be decomposed into the product of two PCO's $\mathbb{Y}^{(0|1)}$ as follows 
 \begin{eqnarray}
\label{SM-NA}
\mathbb{Y}^{(0|2)}  = {\mathbb Y}^{(0|1)}_v  \wedge {\mathbb Y}^{(0|1)}_w  + d \Omega^{(-1|2)} \,.
\end{eqnarray}
where $v$ and $w$ are two independent spinors $Det(v, w) = v^\a \epsilon_{\a\b} w^\b \neq 0$.  
Let us analyse the action with the new choice of PCO:
\begin{eqnarray}
\label{SM-NB}
S_{SCS} = \int_{{\cal SM}^{(3|2)}} \Big( A \wedge d A  + \frac23 A\wedge A\wedge A + W^\a W_\a V^3\Big)\wedge 
  {\mathbb Y}^{(0|1)}_v \wedge {\mathbb Y}^{(0|1)}_w \ ,
  \end{eqnarray}
where the $\Omega$-term is dropped by integration by parts. Let us put aside the interaction term, to be discussed later, and let us distribute the two $\mathbb{Y}$'s on the two pieces 
of the action as follows
\begin{eqnarray}
\label{SM-NC}
S^{quad}_{SCS} = \int_{{\cal SM}^{(3|2)}} \Big( A \wedge d A \wedge {\mathbb Y}^{(0|1)}_v  {\mathbb Y}^{(0|1)}_w    + 
W^\a W_\a   {\mathbb Y}^{(0|1)}_v  {\mathbb Y}^{(0|1)}_w \wedge V^3\Big)\,. 
  \end{eqnarray}
Since the PCO's are closed, we can also bring them after each connection term $A^{(1|0)}$ and 
after the spinorial $W^{(0|0)}$ forms as 
\begin{eqnarray}
\label{SM-NC}
S^{quad}_{SCS} = \int_{{\cal SM}^{(3|2)}} \Big( (A {}_\wedge {\mathbb Y}^{(0|1)}_v)  \wedge d (A {}_\wedge {\mathbb Y}^{(0|1)}_w)   + 
(W^\a {}_\wedge {\mathbb Y}^{(0|1)}_v) \wedge (W_\a {}_\wedge {\mathbb Y}^{(0|1)}_w) \wedge V^3\Big) \ ,
\end{eqnarray}
converting the gauge connection to a $(1|1)$ form as 
\begin{eqnarray}
\label{SM-NCA}
A^{(1|0)} \rightarrow  A^{(1|1)}  \equiv A^{(1|0)} {}_\wedge {\mathbb Y}^{(0|1)}_v \,. 
\end{eqnarray}
In the same way, the $(0|0)$-form $W^\a$ is converted into a $(0|1)$-pseudoform. 
Passing from $A^{(1|0)}$, which has a finite number 
of components, to  $A^{(1|1)}$, we have moved to an infinite dimensional space. 
Therefore, we now take into account the more
generic action  
\begin{equation}\label{CSA}
S_{SCS} = \int_{{\cal SM}^{(3|2)}}  \Big( A^{(1|1)} \wedge d A^{(1|1)}  + 
W^{(0|1), \alpha} \epsilon_{\a\b} \wedge W^{(0|1), \beta} \wedge V^3  \Big)   \,.
\end{equation}
The wedge product is taken in the space of pseudoforms and we use the convention 
that two $(0|1)$-forms must be multiplied with the wedge product. 

Now, we relax the condition (\ref{SM-NCA}) and we consider the action in terms of the 
picture one fields $A^{(1|1)}$. That opens up to a completely new theory with several implications 
still to be fully explored. The strategy is to relax the factorization properties and consider all 
possible terms into a CS-type of action. First, we briefly review the result of our previous paper \cite{Cremonini:2019aao} 
and then we build the BV formalism for this new theory. 

Decomposing the pseudoform $\displaystyle A^{(1|1)} = A_0 + A_1 + A_2 + A_3 $, where the subscript denotes the number of $dx$'s in the expression, we have:
\begin{eqnarray}\nonumber
	\label{TLAAA} A_0 &=& \sum_{p=0}^{\infty} A_{\alpha \beta}^{(p)} ( d \theta^\alpha )^{p+1} \delta^{(p)} ( d \theta^\beta ) \ , \\\nonumber
	\label{TLAAB} A_1 &=& \sum_{p=0}^{\infty} dx^m A_{m \alpha \beta}^{(p)} ( d \theta^\alpha )^{p} \delta^{(p)} ( d \theta^\beta ) \ , \\\nonumber
	\label{TLAAC} A_2 &=& \sum_{p=0}^{\infty} dx^m dx^n A_{[m n] \alpha \beta}^{(p)} ( d \theta^\alpha )^{p} \delta^{(p + 1)} ( d \theta^\beta ) \ , \\
	\label{TLAAD} A_3 &=& \sum_{p=0}^{\infty} dx^m dx^n dx^r A_{[m n r] \alpha \beta}^{(p)} ( d \theta^\alpha )^{p} \delta^{(p + 2)} ( d \theta^\beta ) \ .
\end{eqnarray}

We can compute the Lagrangian:
\begin{eqnarray}\nonumber
	\nonumber \mathcal{L}^{(3|2)} &=& A^{(1|1)}dA^{(1|1)} = 
	\\ \nonumber &=& \sum_{p=0}^\infty \left[ - p! (p+1)! A_{\alpha \beta}^{(p)} \left( \partial_{[r} A_{m n] \beta \alpha}^{(p)} - \partial_\beta A_{[m n r] \beta \alpha}^{(p - 1)} + (p+2) \partial_\alpha A_{[m n r] \beta \alpha}^{(p)} \right) + \right. \\\nonumber
	\nonumber & - & p! p! A_{[m \alpha \beta}^{(p)} \left( \partial_r A_{n] \beta \alpha}^{(p)} - \partial_\beta A_{n r] \beta \alpha}^{(p - 1)} + (p+1) \partial_\alpha A_{n r] \beta \alpha}^{(p)} \right) + \\\nonumber
	\label{TLC} & - & p! (p+1)! A_{[m n \alpha \beta}^{(p)} \left( \partial_{r]} A_{\beta \alpha}^{(p)} - \partial_\beta A_{r] \beta \alpha}^{(p)} + (p+1) \partial_\alpha A_{r] \beta \alpha}^{(p + 1)} \right) + \\
	\nonumber & - & p! (p+2)! A_{[m n r] \alpha \beta}^{(p)} \left( - \partial_\beta A_{\beta \alpha}^{(p)} + (p+1) \partial_\alpha A_{\beta \alpha}^{(p+1)} \right) \Big] dx^m \wedge dx^n \wedge dx^r \delta( d \theta^\beta ) \delta ( d \theta^\alpha ) \  . 
\end{eqnarray}
We obtain the equations of motion by varying the action w.r.t. the fields  $A_{\alpha \beta}^{(p)},$ $A_{m \alpha \beta}^{(p)}$, $A_{[m n] \alpha \beta}^{(p)}$ and $A_{[m n r] \alpha \beta}^{(p)}$; the resulting equations are
\begin{eqnarray}\nonumber
	\label{EMAD} - \partial_\beta A_{\beta \alpha}^{(p)} + (p + 1 ) \partial_\alpha A_{\beta \alpha}^{(p+1)} = 0 \ , \ \forall \ p \in \mathbb{N} \ , \\\nonumber
	\label{EMAE} \partial_r A_{\beta \alpha}^{(p)} - \partial_\beta A_{r \beta \alpha}^{(p)} + (p+1) \partial_\alpha A_{r \beta \alpha}^{(p + 1)} &=& 0 \ \forall \ p \in \mathbb{N} \ , \\\nonumber
	\label{EMAF} \partial_{[ r} A_{n ] \beta \alpha}^{(p)} - \partial_\beta A_{[n r] \beta \alpha}^{(p - 1)} + (p+1) \partial_\alpha A_{[n r] \beta \alpha}^{(p)} &=& 0 \ \forall \ p \in \mathbb{N} \ , \\
	\label{EMAG} \partial_{[r} A_{m n] \beta \alpha}^{(p)} - \partial_\beta A_{[m n r] \beta \alpha}^{(p-1)} + \left( p+2 \right) \partial_\alpha A_{[m n r] \beta \alpha}^{(p)} &=& 0 \ \forall \ p \in \mathbb{N} \ .
\end{eqnarray}
where we stress that in (\ref{EMAF}) and in (\ref{EMAG}) if $p=0$ the fields $A_{[n r] \beta \alpha}^{(- 1)}$ and $A_{[m n r] \beta \alpha}^{(-1)}$ are both defined to be zero. The equations of motion for $W^{(0|1),\a}$ implies that 
this is set to zero algebraically, and this is automatically achieved also in the present framework. 

To show that this complicate set of linear equations coincides with the usual CS equations of motion, one needs to remove the infinite redundancy by algebraic (i.e. $\theta$-dependent) gauge transformations. As a result we find that a representative of the cohomology class is
\begin{equation} \label{CSFIELD}
	A^{1|1} = dx^m \theta^\beta \tilde{B}_{m \alpha \beta}^{(0)} (x) \delta^{(0)} ( d \theta^\beta ) \ ,
\end{equation}
and the relative equation of motion is
\begin{equation}
	\partial_{[n} \tilde{B}_{m] \alpha \beta}^{(0)} (x) = 0 \ .
\end{equation}
Remarkably, notice that even if we started from a SCS Lagrangian with an infinite number of fields, we have shown that there is only one physical field. All the other fields are $d$-exact $\theta$-dependent terms.

Moreover we have shown that starting from the free SCS action with a general $A^{1|1}$ pseudoform we obtain the factorisation
\begin{equation}
	A^{(1|1)} = A^{(1|0)} \wedge \mathbb{Y}^{(0|1)} \ , \ \text{ s.t. } \mathbb{Y}^{(0|1)} = 
	\theta^\beta \delta ( d \theta^\beta ) + d \Omega^{(-1|1)} \ .
\end{equation}
Thus we have recovered a factorised form from a non-factorised Lagrangian.

\newcommand{\MM}{\mathfrak M}
\newcommand{\LL}{\mathfrak L}

\subsection{Interaction terms}

We now define an interaction term which can be integrated on a supermanifold. 
Apparently, a problem arises. In order to define an interaction term, we need three gauge fields $A^{(1|1)}$, but the wedge product of three fields vanishes by the anticommutativity of the three Dirac delta functions of $d\theta_1$ or 
$d\theta_2$. 

In \cite{Cremonini:2019aao} we propose an action where $Z$ is inserted into the product of three gauge fields, however, as discussed in the paper, we have to consider all the possible places where to put the PCO. Therefore, following \cite{Catenacci-Grassi-Noja,Erler:2013xta} (see also the proceeding \cite{Sachs:2019gue}), we are led to define the $2$-$product$ $with$ $picture$ $degree$ -$1$ as
	\begin{align} \label{APA}\nonumber 
		\MM_2 : \Omega^{(1|1)} \times \Omega^{(1|1)} & \to \Omega^{(2|1)} \\
		(A,A) & \mapsto \MM_2 (A,A) = \frac{1}{3} \Big[ Z_v ( A \wedge A ) + Z_v (A) \wedge A + A \wedge Z_v (A) \Big] \ .
	\end{align}
Observe that this product has form degree 0, i.e. it does not change the form number and it decreases the 
picture number by one. The products of the various fields involves also the matrix multiplication of 
the generator in the adjoint representation. 

In an analogous way, we can define a product with form degree $-1$ as
	\begin{align} 
		& \tilde{m}_2^{(-1)} : \Omega^{(1|1)} \times \Omega^{(1|1)}  \to \Omega^{(1|1)} \\
		\label{APD} (A,A) & \mapsto \tilde{m}_2^{(-1)} (A,A) = \frac{1}{3} \left[ -i \Theta (\iota_v) ( A \wedge A ) -i \Theta (\iota_v) (A) \wedge A - (-1)^{|A|} A \wedge i \Theta (\iota_v) (A) \right] \ .\nonumber 
	\end{align}
	This product is needed as an intermediate product to define higher order product as explained in 
	the literature. It maps the integral forms into the space of inverse forms which do not have a physical 
	interpretation (it is analogous to the so called \emph{Large Hilbert Space}). 
	From the definition \eqref{APA}, it follows that
	\begin{equation} \label{APE}
		\MM_2= [ d , \tilde{m}_2^{(-1)} ] \ ,
	\end{equation}
	where $ [ \cdot , \cdot ] $ denotes as usual the graded commutator. 

We can compute the first interaction term of the Lagrangian $\displaystyle A^{(1|1)} \wedge \MM_2 \left( A^{(1|1)} , A^{(1|1)} \right) $, which explicitly reads
\begin{align} 
	\mathcal{L}^{(3|2)}_{INT} = & 2 tr \left\lbrace \sum_{p,q=0}^\infty (-1)^p p! q! q! \left\lbrace (q+1) \left[ A_{m n \alpha \beta}^{(q)} , A_{\beta \alpha}^{(q)} \right] + A_{m \alpha \beta}^{(q)} A_{n \beta \alpha}^{(q)} \right\rbrace \cdot \right. \\
	\label{ITC} & \left. \left\lbrace \left( \frac{v^\alpha}{v^\beta} \right)^p \left( \frac{v^\alpha}{v^\beta} \partial_\alpha + \partial_\beta \right) A_{r \alpha \beta}^{(p)} + \left( \frac{v^\beta}{v^\alpha} \right)^p \left( \frac{v^\beta}{v^\alpha} \partial_\beta + \partial_\alpha \right) A_{r \beta \alpha}^{(p)} \right\rbrace \right\rbrace \epsilon^{mnr} \epsilon^{\alpha \beta} d^3 x \delta^2 \left( d \theta \right) \ .
\end{align}
Notice that the interaction term depends on the constant vector $v^\alpha$ through $\displaystyle \frac{v^1}{v^2}$, namely their relative phase. That resembles the usual frame dependence of Superstring Field Theory actions.

\subsection{Gauge Invariance and the Emergence of the $A_\infty$-Algebra}

The non-associativity of the product $\MM_2$ breaks the gauge invariance of Chern-Simons action, furthermore the algebra of gauge transformations does not close. To overcome these two problems, we need additional 
terms in the action and we need to change the gauge transformations.

We now proceed by constructing explicitly the first multiproduct of the $A_\infty$-algebra.
Let us consider the action discussed so far:
\begin{equation} \label{TPM3A}
	S_A = \int_{{\mathcal SM}^{(3|2)}} {\rm Tr} \left(
	\frac{1}{2}  A \wedge dA + \frac{1}{3} A \wedge \MM_2\left( A , A \right) \right) \ .
\end{equation}
We will assume that the gauge field $A$ is a $(1|1)$-pseudo form and we neglect the $WW$ term 
for the moment (then we denote this part of the action as $S_A$). 
Assuming the cyclicity of the trace, we can compute the field strength from the 
variation of 
  \eqref{TPM3A}
\begin{equation} \label{TPM3B}
	F = dA + \MM_2 ( A , A ) \ .
\end{equation}
The field strength is a $(2|1)$-form as  $dA$, indeed $\MM_2$ consistently reduces the picture of by one.  
Upon applying the exterior derivative $d$, which is a derivation of $\MM_2$ (since $\left[ Z_v , d \right] = 0$), 
we get
\begin{equation} \label{TPM3C}
	d F = d \MM_2 ( A , A ) = \MM_2 ( dA , A ) - \MM_2( A , dA ) \ .
\end{equation}
 We can now use \eqref{TPM3B} to substitute the expression for $dA$  in \eqref{TPM3C} and we get
\begin{equation} \label{TPM3D}
	d F = \MM_2 ( F , A ) - \MM_2 ( A , F ) -  
	\MM_2 \Big( \MM_2 \left( A , A \right) , A \Big) + \MM_2 \Big( A , \MM_2 \left( A , A \right) \Big)  \ ,
\end{equation}
where, as expected, it appears the extra term given by the associator of $\MM_2$. 
In order to get rid of this term, we add an extra term to the action such that
\begin{eqnarray} \label{TPM3DA}
	F' &=& dA + \MM_2 ( A , A ) + \MM_3 ( A , A , A ) = F + \MM_3 ( A , A , A ) \,, \nonumber \\
	d F'  &=&  \MM_2 ( F , A ) - \MM_2 ( A , F ) - \MM_3(dA,A,A) + \MM_3(A, dA,A) - \MM_3(A,A,dA)  \ ,
\end{eqnarray}
which implies that 
\begin{equation} \label{TPM3E}
	d \MM_3 \left( A , A , A \right) + \MM_3 \left( dA , A , A \right) - \MM_3 \left( A , dA , A \right) + 
	\MM_3 \left( A , A , dA \right)  $$ $$ - \MM_2\left( \MM_2\left( A , A \right) , A \right) + \MM_2 
	\left( A , \MM_2\left( A , A \right) \right) = 0 \ .
\end{equation}
This is the third $A_\infty$-relation (for example see \cite{Stasheff,Kajiura:2003ax,Keller,Aspinwall}). 
Notice that when we apply $d$ to $\MM_3$, since, a priori it is not a derivation of $\MM_3$, we also need the 
three additional terms in the first line of (\ref{TPM3E}). The explicit form of $\MM_3$ is given in \cite{Cremonini:2019aao} and we do not repeat here its construction: it is based on $\tilde{m}_2$ which is built from $\Theta(\iota_v)$ instead of $Z$. 

Clearly the Bianchi identity for $F'$ does not hold, in particular from \eqref{TPM3DA} we have
\begin{eqnarray}
	\nonumber d F' &= & -\MM_2 ( F , A ) + \MM_2( A , F ) - \MM_3(dA,A,A) + \MM_3(A, dA,A) - \MM_3(A,A,dA) 
	 \\ \nonumber 
	&= &  - \MM_2\Big( F' - \MM_3(A , A , A ) , A \Big) + \MM_2 \Big( A , F' - \MM_3( A , A , A ) \Big) 
	\\\nonumber 
	&- & 
	 \MM_3\Big(F' - \MM_2(A,A) - \MM_3(A,A,A),A,A\Big) + \MM_3\Big(A, F' -  \MM_2(A,A) - \MM_3(A,A,A),A\Big) 
	\\\nonumber 
	&- & 
	  \MM_3\Big(A,A,F' -  \MM_2(A,A) - \MM_3(A,A,A)\Big) 
	 \\\nonumber 
	& = & \MM_2 ( F' , A ) - \MM_2( A , F' ) - \MM_3(F' ,A,A) + \MM_3(A, F',A) - \MM_3(A,A,F') 
	 \\\nonumber
	& - &
	 \MM_2 \Big( \MM_3 ( A , A , A ) , A ) + \MM_2 \Big( A , \MM_3 ( A , A , A )\Big) 
	 +\MM_3\Big(\MM_2(A,A),A,A\Big) 
	 \\
	& - &
	  \MM_3\Big(A, \MM_2(A,A), A\Big) + \MM_3\Big(A, A, \MM_2(A,A)\Big) + {\cal O}(A^5) \ ,
\end{eqnarray}
where we have neglected the 5-gauge field terms ${\cal O}(A^5)$. 
 Expressing the field strength $F$ in terms of the corrected one $F' = dA + \MM_2(A,A) + \MM_3(A,A,A) + {\cal O}(A^4)$ the equation can be rewritten as 
\begin{eqnarray}
	 \nonumber d F' 
		&-& \MM_2 ( F' , A ) + \MM_2( A , F' ) + \MM_3(F' ,A,A) - \MM_3(A, F',A) + \MM_3(A,A,F') 
	 \\\nonumber
	& = &
	 - \MM_2 \Big( \MM_3 ( A , A , A ) , A ) + \MM_2 \Big( A , \MM_3 ( A , A , A )\Big) 
	 -\MM_3\Big(\MM_2(A,A),A,A\Big) 
	 \\
	& - &
	  \MM_3\Big(A, \MM_2(A,A), A\Big) + \MM_3\Big(A, A, \MM_2(A,A)\Big) + {\cal O}(A^5) \ ,
\end{eqnarray}
where in the first line we have the Bianchi indentities, broken by 
the right-hand side of the equation, which contains 4 gauge field terms expressing 
the non-associativity of the $\MM_2$ and $\MM_3$ products. 
Again, we have an extra term breaking the Bianchi identity. By following the prescription described above, we add to the action an extra term, $\MM_4(A,A,A,A)$ for the field strength and correspondently for the 
action. Proceeding in the same way, we have a new action of the 
form 
\begin{eqnarray}
\label{novaA}
S_A = \int_{{\mathcal SM}^{(3|2)}} {\rm Tr} \left(
	\frac{1}{2}  A dA + \sum_{n=2}^\infty \frac{1}{n+1} A \wedge \MM_n\left( A , \dots, A \right) \right) \ .
\end{eqnarray}
yielding the equations of motion
\begin{eqnarray}
\label{novaB}
d A + \sum_{n} \MM_n(A, \dots,A) =0\,. 
\end{eqnarray}
 consistent with the Bianchi identities because of the $A_\infty$ relations among the various multiproducts 
 $\MM_n(A, \dots, A)$. Notice that according to \cite{Cremonini:2019aao} every multiproduct reduces the form degree 
 from $n$ (the total degree of the product of $n$ gauge fields $A$) to 2 and the 
 picture number from $n$ (the total picture of a formal product of $n$ gauge fields) down to 1 as required for building an integral form.  

Let us now study the gauge symmetry. Previously we have seen that, since the product $m_2^{(-1)}$ is not associative, the gauge algebra does not close. We now show that, in order to close the algebra, we have to modify the gauge transformation law by introducing multiproducts induced by the $A_\infty$ algebra discussed so far, but 
then it emerges that one necessarily needs the BV formalism to deal with the gauge symmetries. 
We derive the BRST symmetry by shifting the gauge field $A$ with $A + c$. Notice that we trade the form number with the ghost number, but we have not changed the picture number, since 
both $A$ and $c$ have the same picture. The second shift is $d$ into $d + s$, introducing the BRST differential.  
Then we impose the equation 
\begin{eqnarray}
\label{RUSA}
(d + s) (A + c) + \MM_2(A+c, A+c) + \MM_3(A+c,A+c,A+c) + {\mathcal O}(A^4, \dots, c^4) = F' \ ,
\end{eqnarray}
from which we get the following relations 
\begin{eqnarray}
\label{RUSB}
F' &=& d A + \MM_2(A,A) + \MM_3(A,A,A) +  {\mathcal O}(A^4)\,, \nonumber \\
0  &=& s A + d c + \MM_2(A,c) + \MM_2(c,A) + \MM_3(A,A,c) + \MM_3(A,c,A) + \MM_3(c,A,A) 
+  {\mathcal O}(A^3)\,, \nonumber \\
0 & =& s c +  \MM_2(c,c) + \MM_3(A,c,c) + \MM_3(c,A,c) + \MM_3(c,c,A) + {\mathcal O}(A^2) \,, \nonumber \\
0 &=& \MM_3(c,c,c) + {\mathcal O}(A)\,. 
\end{eqnarray}
The last equation is consistent with the definition of the $\MM_3$ product since it decreases the 
form number, but $c$ has form degree equal to zero. The first line reproduces the definition of $F'$ given in (\ref{TPM3DA}). 
The second line gives the gauge transformation of the gauge field 
\begin{eqnarray}
\label{RUSC}
s A  = - d c -  \LL_2(c,A) - \frac12 \LL_3(c,A,A)
+  {\mathcal O}(A^3)\,, 
\end{eqnarray}
where the Lie-algebra-like symbols (for anticommuting quantities) 
\begin{eqnarray}
\label{RUSD}
\LL_2(A,c) &=& \MM_2(A,c) + \MM_2(c,A)\,, ~~~~\nonumber \\
\LL_3(A,c,c) &=& 2 \MM_3(A,c,c) + 2 \MM_3(c,A,c) + 2 \MM_3(c,c,A) \,, \nonumber \\
\LL_2(X,Y) &=& \MM_2(X,Y) - (-1)^{|X||Y|} \MM_2(Y,X) \,, \nonumber \\
\LL_3(X_1,X_2,X_3) &=& \MM_3 (X_1,X_2,X_3) - (-1)^{|X_2||X_3|} \MM_3 (X_1,X_3,X_2) \nonumber \\ && + \MM_3 (X_2,X_3,X_1) - (-1)^{|X_3||X_1|} \MM_3 (X_2,X_1,X_3) \nonumber \\ && + \MM_3 (X_3,X_1,X_2) - (-1)^{|X_1||X_2|} \MM_3 (X_3,X_2,X_1) \,, 
\end{eqnarray}
have been introduced. Note that $\LL_2(c,c) = 2 \, \MM_2(c,c)$. In addition, $\LL_2(A,c)$ has 
form number = +1, ghost number =+1, and picture = +1. $\LL_3(A,c,c)$ has form number = 0, 
ghost number = +2 and picture = +1. 
The third line of (\ref{RUSB}) gives us the BRST transformation of the ghost field 
\begin{eqnarray}
\label{RUSE}
s c = - \frac12 \LL_2(c,c) - \frac12 \LL_3(A,c,c) + {\mathcal O}(A^2) \ .
\end{eqnarray}
In order to study the nilpotency of $s$, it is useful to verify the compatibility 
of the multiproducts with $s$. For example, starting from the second $A_\infty$ relation 
we have 
\begin{eqnarray}
\label{RUSF}
(s + d) \MM_2(A+c, A+c) - \MM_2((s+d)(A+c), A+c) + \MM_2(A+c, (s+d)(A+c)) = 0\,, 
\end{eqnarray}
which implies the following relations
\begin{eqnarray}
\label{RUSG}
&&s \MM_2(A, A) - \MM_2( sA, A) + \MM_2(A, s A) = 0\,, \nonumber \\
&&s \LL_2(A, c) - \LL_2( s A, c) - \LL_2(s c, A) = 0\,, \nonumber \\
&&s \LL_2(c,c) - \LL_2(sc ,c) + \LL_2(c, s c) =0\,,
\end{eqnarray}
where $\LL_2( s A, c) = \MM_2(s A, c) - \MM_2(c, s A)$ since $c$ is fermionic and $s A$ is bosonic. This implies 
that $\LL_2(s A, c) = - \LL_2(c, s A)$ and $\LL_2(s c, A) = - \LL_2(A, s c)$, leading to 
\begin{eqnarray}
\label{RUSGa}
s \LL_2(A, c) = \LL_2( s A, c) - \LL_2(A, s c)\,, 
\end{eqnarray}
which expressed the Leibniz rule of $s$ with respect to $\LL_2$. 

Let us study the nilpotency of $s$ on the ghost field. 
Acting with $s$ on (\ref{RUSE}), we get
\begin{eqnarray}
\label{RUSH}
s^2 c &=& -\frac12 s \, \LL_2(c,c) - \frac12 s \LL_3(A,c,c) + {\cal O}(A^2) \nonumber \\
&=&  -\LL_2(sc ,c)  - \frac12 \LL_3(s A,c,c) + \LL_3(A ,s c,c) - \LL \left( A,c,sc \right) + {\cal O}(A^2) \ .
\end{eqnarray}
Inserting the BRST transformations of $c$, using the $A_\infty$ relations,  it yields
\begin{eqnarray}
\label{RUSI}
s \Big( s c - \frac14 \LL_4(A,A,c,c) \Big)
&=&  \frac13 \LL_4(F, c,c,c) + {\cal O}(A^2) \,, 
\end{eqnarray}
where on the left-hand side we have reabsorbed $\LL_4(A,A,c,c)$ in the definition 
of the BRST transformation of the ghost $c$, and on the right-hand side we finally found that 
the algebra is not closed, but it closes on the field strength $F$ of the gauge field $A$. This is 
crucial, since this implies that we need the antifield formalism to deal with it since the field strength 
is just the variation of the antifield $A^*$ of the gauge field. As it will be shown in the last section it 
is precisely the integral form formalism that gives the correct quantum number for $A^*$. It should be 
a $(2|1)$-form.

\section{BV action}

\subsection{Picture-0 Gauge Fields}


To build the \emph{BV action}, we need to include into the previous action the antifields as the generators of the BRST transformations (the superspace formulation has been studied in \cite{Grassi:2004tv}); at the moment we assume these antifields to be superforms. For example, the term corresponding to the BRST transformation of the gauge field reads
\begin{equation}\label{P0GFB}
	\int_{\mathcal{SM}^{(3|2)}} {\rm Tr}
	\left[ A^{*(2|0)} \wedge \nabla c^{(0|0)} \right] \wedge \mathbb{Y}^{(0|2)} \ ,
\end{equation}
where $A^{*(2|0)}$ is the $(2|0)$-superform
\begin
{equation}\label{P0GFC}
	A^* = V^a V^b A^*_{a b} + V^a \psi^\alpha A^*_{a \alpha} + \psi^\alpha \psi^\beta A^*_{\alpha \beta} \ .
\end{equation}
The PCO $\mathbb{Y}^{(0|2)}$ is used to convert it into an integral form. 
Comparing with the component formalism, 
we see that there are too many independent components in $A^*$ and, therefore, we need  a
convenient constraint to reduce them systematically. Hence, we set
\begin{equation}\label{P0GFD}
	\nabla A^*=0\,, ~~~~ A^*_{\alpha \beta} = 0 \ \implies \ 
	A^* = V^a V^b A^*_{a b} + V^a \psi^\alpha \gamma_{a \alpha \beta} W^{* \beta} \ ,
\end{equation}
as for the field strength $F^{(2|0)}$ (see eq. (\ref{SM-A})). $A^*_{ab}$ serves as the antifield for the gauge field, while $W^{*\alpha}$ for the gaugino. The vanishing of the covariant derivative of $A^*$ is needed to require the closure of the $0$-picture factor of (\ref{P0GFB}). Here, this choice seems to be tailored, then the quest for a more natural setting. Nevertheless, by taking the susy PCO, i.e. $\displaystyle \mathbb{Y}^{(0|2)} = V^a V^b \epsilon_{a b c} \iota_\alpha \gamma^{c \alpha \beta} \iota_\beta \delta^2 \left( \psi \right) $, eq.~\eqref{P0GFB} reads
\begin{equation}\label{P0GFE}
	\int_{\mathcal{SM}^{(3|2)}} {\rm Tr}\left[ \left( V^a V^b A^*_{a b} + V^a \psi^\alpha \gamma_{a \alpha \beta} W^{* \beta} \right) \wedge \left( V^c \nabla_c c + \psi^\gamma \nabla_\gamma c \right) \right] \wedge \mathbb{Y}^{(0|2)}_{susy} = \int_{x,\theta} {\rm Tr}(W^{*\alpha} \nabla_\alpha c) \ .
\end{equation}
This result matches with the superspace CS BV-action. 

The integral form formalism provides a more natural way to obtain the correct BV terms. We observe 
that the {\it Serre's} dual to a $(1|0)$-superform is a $(2|2)$ integral form (see also \cite{Castellani:2015ata}):
\begin{equation}\label{P0GFF}
	A^{*(2|2)} = V^a V^b \epsilon_{a b c} A^{*c} \delta^2 \left( \psi \right) + V^a V^b V^c \epsilon_{a b c} W^{*\alpha} \iota_\alpha \delta^2 \left( \psi \right) \ .
\end{equation}
We do not impose any constraint on $A^{*(2|2)}$, since this integral form already contains the correct number of fields. Analogously, the BRST symmetry of the ghost $c$ is coupled to a $(3|2)$-integral form $c^*$, representing its antifield:
\begin{equation}\label{P0GFG}
	c^{*(3|2)} = c^* V^a V^b V^c \epsilon_{a b c} \delta^2 \left( \psi \right)  \ .
\end{equation}
Therefore the action reads
\begin{equation}\label{P0GFH}
	S = \int_{\mathcal{SM}^{(3|2)}} {\rm Tr}\left[ \left( A dA + \frac{2}{3} A^3 + \frac{1}{2} W^2 V^3\right) \wedge \mathbb{Y} + A^* \nabla c + \frac{1}{2}  c^* \left[ c , c \right]\right] = S_{SCS} + S_{BV} \ .
\end{equation}
To compare \eqref{P0GFH} with the component or with the superspace actions, it is convenient to rewrite it into a factorised form ${\mathcal L}^{(3|0)} \wedge {\mathbb Y}^{(0|2)}$. 
For that we have
\begin{eqnarray}\label{P0GFI}
	{\rm Tr} \left(A^* \nabla c +  \frac12 c^* \left[ c , c \right] \right) &=& 
	 Z {\rm Tr} \left( A^* \nabla c +  \frac12 c^* \left[ c , c \right] \right) \wedge \mathbb{Y}\nonumber \\ 
	&=&{\rm Tr} \left[ Z \left( A^* \nabla c \right) + Z \left( \frac12 c^* \left[ c , c \right] \right) \right] \wedge \mathbb{Y} = \mathcal{L}_{BV} \wedge \mathbb{Y} \ ,
\end{eqnarray}
where the formal inverse $Z$ of $\mathbb{Y}$ and the linearity have been used\footnote{Notice that by $Z$ we mean the product of two PCO $Z_v$ and $Z_w$, where $v$ and $w$ represent two independent directions.}. It follows that
\begin{eqnarray}\label{P0GFJ}
	d \mathcal{L}_{BV} &=& d {\rm Tr} \left[ Z \left( A^* \nabla c \right) + Z \left(  \frac12 c^* \left[ c , c \right]\right) \right] = d \, Z \, {\rm Tr} \left( A^* \nabla c \right) + d \, Z \, {\rm Tr} \left(  \frac12 c^* \left[ c , c \right] \right) 
	\nonumber \\&=& Z {\rm Tr}  \left[ d \left( A^* \nabla c \right) + d \left( \frac12 c^* \left[ c , c \right] \right) \right] = 0 \ ,
\end{eqnarray}
using $\displaystyle \left[ d , Z \right] = 0$ and noting that $d$ acts on a top integral form. The closure of the Lagrangian suggests that we do not need any other terms, in particular we do not need any further 
antifield since all the needed dof's are already present. 
To verify this, we compute \eqref{P0GFI} for two choices of PCO, namely the supersymmetric PCO and the component PCO. Let us start from the supersymmetric case. It is easy to verify that
\begin{eqnarray}\label{P0GFK}
	{\rm Tr} A^* \nabla c &=& {\rm Tr}\Big[  
	\left[ V^a V^b \epsilon_{a b c} A^{*c} \delta^2 \left( \psi \right) + V^a V^b V^c \epsilon_{a b c} W^{*\alpha} \iota_\alpha \delta^2 \left( \psi \right) \right] \wedge \left[ V^d \nabla_d c + \psi^\gamma \nabla_\gamma c \right] \Big] \nonumber \\ &=& 
{\rm Tr}\Big[\left[ A^{*a} \psi \gamma_a \psi + V^c W^{*} \gamma_c \psi \right] \wedge \left[ V^d \nabla_d c + \psi^\gamma \nabla_\gamma c \right]  \Big]\wedge \mathbb{Y}_{susy}^{(0|2)} \nonumber \\ &=& 
{\rm Tr}\left[ A^{*a} \psi \gamma_a \psi V^d \nabla_d c + V^c W^{*} \gamma_c \psi \psi^\gamma \nabla_\gamma c \right] \wedge \mathbb{Y}_{susy}^{(0|2)} \ .
\end{eqnarray}
We see that the first term contains the antifield relative to the gauge field $A^a$, while the second term contains the antifield of the gaugino. Eq. \eqref{P0GFK} coincides with the result in \eqref{P0GFE} up to a rescaling: recall that $\displaystyle \nabla_a = \frac{1}{2} \gamma_a^{\alpha \beta} \left\lbrace \nabla_\alpha , \nabla_\beta \right\rbrace $, then
\begin{equation}\label{P0GFL}
	{\rm Tr}\left[ A^{*a} \psi \gamma_a \psi V^d \nabla_d c + V^c W^{*} \gamma_c \psi \psi^\gamma \nabla_\gamma c \right] \wedge \mathbb{Y}_{susy}^{(0|2)} $$$$= 
	{\rm Tr}
	\left[ A^{*a} \psi \gamma_a \psi V^b \gamma_b^{\alpha \beta} \nabla_\alpha \nabla_\beta c + V^c W^{*} \gamma_c \psi \psi^\beta \nabla_\beta c \right] \wedge \mathbb{Y}_{susy}^{(0|2)} $$ $$ =
	{\rm Tr}\Big[ V^c \left[ - \nabla_\alpha A^{*a} \psi \gamma_a \psi \gamma_c^{\alpha \beta} + W^{*} \gamma_c \psi \psi^\beta \right] \nabla_\beta c\Big] \wedge \mathbb{Y}_{susy}^{(0|2)} \ ,
\end{equation}
up to exact terms.

The same analysis can be repeated for the component PCO, namely $\displaystyle \theta^2 \delta^2 \left( d \theta \right)$:
\begin{eqnarray}\label{P0GFM}
	{\rm  Tr}A^* \nabla c &=& V^a V^b V^c \epsilon_{a b c} 
	{\rm Tr}\left[ A^{*d} \nabla_d c - W^{*\alpha} \nabla_\alpha c \right] \delta^2 \left( \psi \right) \nonumber 
	\\ &=& dx^a dx^b dx^c \epsilon_{abc}{\rm Tr} \left[ A^{*d} \nabla_d c - W^{*\alpha} \nabla_\alpha c \right] \delta^2 \left( d \theta \right) \ ,
\end{eqnarray}
passing from the vielbein basis to the component one given by $dx$'s and $d \theta$'s. We can now apply the PCO $Z$ and by recalling that $Z \left(f \delta^2 \left( d \theta \right) \right)= \partial_{\theta^1} \partial_{\theta^2} f$ we get
\begin{equation}\label{P0GFN}
	Z \left[ dx^a dx^b dx^c \epsilon_{abc} {\rm Tr}\left( A^{*d} \nabla_d c - W^{*\alpha} \nabla_\alpha c \right) \delta^2 \left( d \theta \right) \right] 
	$$ $$= dx^a dx^b dx^c \epsilon_{abc} {\rm Tr}\Big(
	 \partial_\theta^2 \left( A^{*d} \nabla_d c - W^{*\alpha} \nabla_\alpha c \right) \Big) = dx^a dx^b dx^c \epsilon_{abc} {\rm Tr} \Big( 
	 \partial_\theta^2 \left[ \left( - \nabla_\alpha A^{*d} \gamma_d^{\alpha \beta} - W^{*\beta} \right) \nabla_\beta c  \right]\Big) \ . 
\end{equation} 
Again the gaugino antifield emerges and it couples correctly to the BRST variation of the fields. 

\subsection{Picture-1 Gauge Fields}

We now want to study the BV formalism in the context of picture 1 fields, namely pseudoforms. Let us start by considering the action \eqref{novaA}. We can substitute the gauge field $A^{(1|1)}$ with a form $\mathcal{A}$ with general form number and picture number 1:
\begin{equation}\label{P1GFA}
	S_{SCS-BV} = \int_{\mathcal{SM}^{(3|2)}} {\rm Tr} \left[ \frac{1}{2} \mathcal{A} \wedge d \mathcal{A} + \mathcal{A}  \wedge\sum_{i=2}^{\infty} \frac{1}{i+1} \MM_i \left( \mathcal{A} , \mathcal{A}, \ldots \right) \right] \ .
\end{equation}
The gauge transformations are obtained by applying the rules described in the previous sections: we replace  the exterior derivative $d$ with $d+s$ in the e.o.m., and the $(1|1)$ gauge field $A^{(1|1)}$ 
with a sum of $(p|1)$ forms, $p \in \mathbb{Z},$\footnote{Notice that this formula is analogous to the string field in bosonic or super string field theory where the form number is replaced by the ghost number and the physical fields are those with ghost number equal to 1; the components with non-positive form numbers are interpreted as the ghosts and ghosts-for-ghosts of several generations, those with form number greater than 1 are interpreted as the antifields of the gauge field and of the entire set of ghosts.} namely
\begin{equation}\label{P1GFB}
	d \to d + s \ , \ A^{(1|1)} \to \mathcal{A} = \sum_{p=-\infty}^{\infty} A^{(p|1)} \ \implies $$ $$ \implies \left( d + s \right) \mathcal{A} + \sum_{i=2}^{\infty} \MM_i \left( \mathcal{A} , \mathcal{A} , \ldots \right) = 0 \ .
\end{equation}
The action written in these terms contains an infinite number of terms with all powers of any $A^{(p|1)}$. 

The BV terms look quite redundant compared to the previous results with a $2$-form and a $3$-form only; 
at picture number 1, we firstly need $A^{(2|1)}$ and $A^{(3|1)}$. 
These two fields are the natural antifields for $A^{(1|1)}$ and for $A^{(0|1)} \equiv c^{(0|1)}$ respectively. Serre's duality at picture 1 involves only picture-one forms. Then we denote them as:
\begin{equation}\label{P1GFC}
	A^{(2|1)} \equiv A^{*(2|1)} \ , \ A^{(3|1)} \equiv c^{*(3|1)} \ .
\end{equation}
Notice that a priori we may have pseudoforms with different form numbers as well, for example in the action there can be a term as
\begin{equation}\label{P1GFD}
	A^{(-1|1)} \wedge \MM_2 \left( A^{*(2|1)} , A^{*(2|1)} \right) \ .
\end{equation}
In order to justify these structures, we proceed into two steps. First, we show that there is always a field redefinition 
which can re-express the entire set of forms into those with $0\leq p\leq3$, but 
the BV action and the BV symplectic form are not compatible with that field 
redefinition (see \cite{Gaberdiel:1997ia} for a detailed description of compatible 
field redefinition for $A_\infty$ BV string field theory action\footnote{In  
\cite{Gaberdiel:1997ia}, the authors show that in the case of tensor construction, the 
\virgolette gauge group" d.o.f. -- which in principle could be described by a $A_\infty$ algebra with an even 
symplectic form -- can be suitably redefined, under adequate hypotheses, and the only non-trivial structure is an associative 
algebra.}). In particular, we show that this field redefinition implies non-trivial constraints on the set of fields.

First, we show that we need to include all possible BV fields, in particular those with negative and 
positive (greater than three) form number fields by studying the structure of multiproducts.  For that 
we observe that we can rewrite the fields by
starting from the expansion of a general picture-1 form on a basis of forms:
\begin{equation}\label{SEA}
	\mathcal{A}^{(\bullet |1)} = \sum_{p=-\infty}^{\infty} A^{(p|1)} = \sum_{i=1,2} \sum_{n=0}^{3} \sum_{q=-\infty}^\infty A^i_{n,q} e_{n,i}^{q |1} \ \ \ \in \ \bigoplus_{r=-\infty}^\infty \Omega^{(r|1)} \ .
\end{equation}
This notation means that the bosonic form number is $n$, the fermionic form number is $q$ while the index $i$ indicates the argument of the delta function, i.e. $d \theta_i$. In particular, we have
\begin{equation}\label{SEB}
	e_{n,i}^{q |1} = \underbrace{ dx \wedge \ldots \wedge dx }_{n} \sum_{p=0}^{\infty} \left( d \theta_j \right)^{p} \left( \iota_i \right)^p \begin{cases}
		\left( d \theta_j \right)^q \delta \left( d \theta_i \right) \ , \ \text{ if } \ q \geq 0 \ , \\
		\iota_i^q \delta \left( d \theta_i \right) \ , \ \ \ \ \ \ \ \text{ if } \ q \leq 0 \ ,
	\end{cases} \ i \neq j \ .
\end{equation}
We single out a box defined by the conditions $0 \leq n \leq 3$ and $0 \leq n + q \leq 3$:
\begin{equation}\label{SEC}
	e_{n,i}^{-q|1} \ \ \ \ \ \ \begin{matrix}
		e_{0,i}^{0 |1} \ & \ e_{0,i}^{1 |1} \ & \ e_{0,i}^{2 |1} \ & \ e_{0,i}^{3 |1} \\
		e_{1,i}^{-1 |1} \ & \ e_{1,i}^{0 |1} \ & \ e_{1,i}^{1 |1} \ & \ e_{1,i}^{2 |1} \\
		e_{2,i}^{-2 |1} \ & \ e_{2,i}^{-1 |1} \ & \ e_{2,i}^{0 |1} \ & \ e_{2,i}^{1 |1} \\
		e_{3,i}^{-3 |1} \ & \ e_{3,i}^{-2 |1} \ & \ e_{3,i}^{-1 |1} \ & \ e_{3,i}^{0 |1}
	\end{matrix} \ \ \ \ \ \ e_{n,i}^{q|1} \ \ \ , q \geq 0 \ \ \ .
\end{equation}
As a consequence of \eqref{SEB} we can write every element of the full basis as elements of the diagonal of the cohomological box (we call it \virgolette cohomological box" since the whole cohomology is contained in those spaces only, see e.g. \cite{Catenacci-Grassi-Noja,Cremonini:2019aao}) as follows:
\begin{equation}\label{SED}
	e_{n,i}^{q |1} = \underbrace{ dx \wedge \ldots \wedge dx }_{n} \sum_{p=0}^{\infty} \left( d \theta_j \right)^{p} \left( \iota_i \right)^p \begin{cases}
		\left( d \theta_j \right)^q \delta \left( d \theta_i \right) = \left( d \theta_j \right)^q e_{n,i}^{0 |1} \ , \ \text{ if } \ q \geq 0 \ , \\
		\iota_i^q \delta \left( d \theta_i \right) = \iota_i^q e_{n,i}^{0 |1} \ , \ \ \ \ \ \ \ \text{ if } \ q \leq 0 \ ,
	\end{cases} \ i \neq j \ .
\end{equation}

The elements of the box in (\ref{SEC}) have a natural interpretation: they provide a basis for
the ghost $c$, the gauge field $A$, the antifield $A^*$ and the antighost $c^*$:
\begin{eqnarray}
	\nonumber c^{(0|1)} &=& A_0^0 +  A_1^{-1} +  A_2^{-2} +  A_3^{-3} \ , \\
	\nonumber A^{(1|1)} &=& A_0^1 +  A_1^0 +  A_2^{-1} +  A_3^{-2} \ , \\
	\nonumber A^{*(2|1)} &=& A_0^{*2} +  A_1^{*1} +  A_2^{*0} +  A_3^{*-1} \ , \\
	\label{BCGA} c^{*(3|1)} &=& c_0^{*3} +  c_1^{*2} +  c_2^{*1} +  c_3^{*0} \ .
\end{eqnarray}

Let us first focus on the gauge field only. In \cite{Cremonini:2019aao} 
we have explicitly shown that, on-shell, the complete tower of fields reduces to the first field appearing in $A_1^0$, namely it has the factorized form
\begin{equation}\label{BCGB}
	A^{(1|1)}_{phys} = dx^a A_a \mathbb{Y}^{(0|1)} \ ,
\end{equation}
where $\mathbb{Y}^{(0|1)}$ denotes as usual the PCO. In particular we observe that \eqref{BCGB} is 
$A_1^0$ in the table above. Moreover, we have shown in \eqref{SED} that, by using carefully the differential operators $d \theta$ and $\iota_\theta$, we can write every element of the basis \eqref{SEC} by using the basis elements $\displaystyle e_{n,i}^{0|1} , n = 0,1,2,3$ only. We could then be induced to consider the BV action to be built by ghost, gauge field, antifield and antighost of the following form:
\begin{equation}\label{BCGC}
	c_0^0 \ , \ A_1^0 \ , \ A^{*0}_2 \ , \ c^{*0}_3 \ .
\end{equation}
This would be analogous to consider the elements of the diagonal of \eqref{SED}. However, this is not the case; indeed, we know that the gauge field is made of four pieces:
\begin{equation}\label{BCGD}
	A^{1|1} = A_0^1 + A_1^0 + A_2^{-1} + A_3^{-2} \ ,
\end{equation}
and this is required by covariance on a generic supermanifold. An analogous argument holds for the antifield and the antighost, 
then we need the whole set of fields in \eqref{BCGA}. Let us consider now the structure of the products of the action \eqref{P1GFA}; a simple power counting shows that we have to take into account not only the fields of \eqref{BCGA}, but also those of the form \eqref{SEA}. Indeed, recall that
\begin{equation}\label{BCGE}
	\MM_2 : \Omega^{(p|1)} \otimes \Omega^{(q|1)} \to \Omega^{(p+q|1)} \ .
\end{equation}
This means we generate all other possible 
sets, even if we started from the fields in \eqref{BCGA}. 
For example, we generate forms of $\Omega^{(6|1)}$ from the 2-products
\begin{equation}\label{BCGF}
	\MM_2 \left( c_0^{*3} , c_0^{*3} \right) \ \in \ \Omega^{(6|1)} \ .
\end{equation}
In particular this term would appear in the action as
\begin{equation}\label{BCGG}
	B_3^{-6} \wedge \MM_2 \left( c_0^{*3} , c_0^{*3} \right) \ ,
\end{equation}
where $B_3^{-6} \in \Omega^{(-3|1)}$. That shows we have to include also the fields with negative form degree, which, in turn,  will enter in the multiproducts themselves.

These arguments show that the definitions of the multiproducts require fields with any form degree. Their structure is given by the geometric data of the supermanifold considered. Hence, the $A_\infty$-structure of the multiproducts and the $L_\infty$-structure of the gauge transformations are the building blocks of an action on a supermanifold with fields with non-zero and non-maximum picture number.

Now, we have to show that there is no non-trivial field redefinition which allows us to restrict the space of fields 
to those of (\ref{BCGA}). For that we start from the BV symplectic form and we rewrite it as follows:
\begin{equation}\label{SEE}
	\omega_{BV} = \left\langle \mathcal{A} , \mathcal{A} \right\rangle = 
	\int_{{\cal SM}} \mathcal{A}^{(\bullet|1)}\wedge \mathcal{A}^{(\bullet|1)} \ ,
\end{equation}
and the integral selects the forms with total form degree equal to three. 
We claim that with the construction described in the previous paragraphs we prove that there is no unconstrained field redefinition establishing the equivalence between the BV theory written using all possible components of $\mathcal{A}$ with any form number, and the BV theory written using the fields of the box (\ref{BCGA}) only. Let us consider a term of \eqref{SEE} made as
\begin{equation}\label{SEF}
	\left\langle A^{(p|1)} , A^{(3-p|1)} \right\rangle = \left\langle A^i_{n,q} e_{n,i}^{q|1} , A^j_{3-n,-q} e_{3-n,i}^{-q|1} \right\rangle = $$ $$ = \int_{\mathcal{SM}} Tr \left\lbrace \sum_{p=0}^\infty (-1)^{|A^{j(p)}_{3-n,-q}|} A^{i,(p)}_{n,q} A^{j,(p)}_{3-n,-q} \left( d \theta_j \right)^{p+q} \left( d \theta_i \right)^{p} \iota_i^p \delta \left( d \theta_i \right) \iota_j^{p+q} \delta \left( d \theta_j \right) \right\rbrace = $$ $$ = \int_{\mathcal{SM}} Tr \left\lbrace \sum_{p=0}^\infty (-1)^{|A^{j(p)}_{3-n,-q}|} A^{i,(p)}_{n,q} A^{j,(p)}_{3-n,-q} (-1)^{q} \left[ \iota_j^q \left( d \theta_j \right)^{p+q} \right] \left( d \theta_i \right)^{p} \iota_i^p \delta \left( d \theta_i \right) \iota_j^{p} \delta \left( d \theta_j \right) \right\rbrace = $$ $$ = \int_{\mathcal{SM}} Tr \left\lbrace \sum_{p=0}^\infty (-1)^{|A^{j(p)}_{3-n,-q}|} A^{i,(p)}_{n,q} A^{j,(p)}_{3-n,-q} \frac{(q+p)!}{p!} \left( d \theta_i \right)^{p} \left( d \theta_j \right)^p \iota_i^p \delta \left( d \theta_i \right) \iota_j^{p} \delta \left( d \theta_j \right) \right\rbrace = $$ $$ = \int_{\mathcal{SM}} Tr \left\lbrace \sum_{p=0}^\infty (-1)^{|A^{j(p)}_{3-n,-q}|} \tilde{A}^{i,(p)}_{n,q} \tilde{A}^{j,(p)}_{3-n,-q} \left( d \theta_i \right)^{p} \left( d \theta_j \right)^p \iota_i^p \delta \left( d \theta_i \right) \iota_j^{p} \delta \left( d \theta_j \right) \right\rbrace = $$ $$ = \left\langle \tilde{A}^i_{n,q} e_{n,i}^{0|1} , \tilde{A}^j_{3-n,-q} e_{3-n,i}^{0|1} \right\rangle = \left\langle \tilde{A}^i_{n,q} \tilde{A}^j_{3-n,-q} e_{n,i}^{0|1} , e_{3-n,i}^{0|1} \right\rangle \ ,
\end{equation}
where the $\sim$ is used to denote the redefinition of the fields in order to absorb the combinatorial coefficients.

To establish the explicit equivalence of the two BV symplectic forms we end up with constraints of the fields.
The BV from \eqref{SEE} with only the fields and antifields $c^{(0|1)} , A^{(1|1)} , A^{*(2|1)} , c^{*(3|1)}$ reads
\begin{equation}\label{SEG}
	\omega_{BV,A} = \left\langle c^{(0|1)} + A^{(1|1)} + A^{*(2|1)} + c^{*(3|1)} , c^{(0|1)} + A^{(1|1)} + A^{*(2|1)} + c^{*(3|1)} \right\rangle = $$ $$ = 2 \left\langle c^{(0|1)} , c^{*(3|1)} \right\rangle + 2 \left\langle A^{(1|1)} , A^{*(2|1)} \right\rangle = $$ $$ = 2 \sum_{n=0}^3 \left[ \left\langle c_n^{(-n|1)} , c_{3-n}^{*(n|1)} \right\rangle + \left\langle A_n^{(1-n|1)} , A_{3-n}^{*(n-1|1)} \right\rangle \right] = $$ $$ = 2 \sum_{n=0}^3 \left[ \left\langle c^i_{n,-n} e_{n,i}^{(-n|1)} , c^{*j}_{3-n,n} e_{3-n,j}^{(n|1)} \right\rangle + \left\langle A^i_{n,1-n} e_{n,i}^{(1-n|1)} , A^{*j}_{3-n,n-1} e_{3-n,j}^{(n-1|1)} \right\rangle \right] = $$ $$ = 2 \sum_{n=0}^3 \left[ \left\langle \tilde{c}^i_{n,-n} e_{n,i}^{(0|1)} , \tilde{c}^{*j}_{3-n,n} e_{3-n,j}^{(0|1)} \right\rangle + \left\langle \tilde{A}^i_{n,1-n} e_{n,i}^{(0|1)} , \tilde{A}^{*j}_{3-n,n-1} e_{3-n,j}^{(0|1)} \right\rangle \right] = $$ $$ = 2 \sum_{n=0}^3 \left[ \left\langle \left( \tilde{c}^i_{n,-n} \tilde{c}^{*j}_{3-n,n}  + \tilde{A}^i_{n,1-n} \tilde{A}^{*j}_{3-n,n-1} \right) e_{n,i}^{(0|1)} , e_{3-n,j}^{(0|1)} \right\rangle \right] = \sum_{n=0}^3 \left[ \left\langle B_n^{ij} e_{n,i}^{(0|1)} , e_{3-n,j}^{(0|1)} \right\rangle \right] \ ,
\end{equation}
where in the last equation we have written the field in brackets in terms of a single field to make the notation easier. Now we can repeat the same calculations for the BV symplectic form built with the the whole set of fields:
\begin{equation}\label{SEH}
	\omega_{BV,\mathcal{A}} = \sum_{p=-\infty}^{\infty} \left\langle A^{(p|1)} , A^{(3-p|1)} \right\rangle = \sum_{n=0}^3 \sum_{p=-\infty}^{\infty} \left\langle A^i_{n,p-n} e_{n,i}^{p-n|1} , A^j_{3-n,n-p} e_{3-n,j}^{n-p|1} \right\rangle = $$ $$ = \sum_{n=0}^3 \sum_{p=-\infty}^{\infty} \left\langle \tilde{A}^i_{n,p-n} e_{n,i}^{0|1} , \tilde{A}^j_{3-n,n-p} e_{3-n,j}^{0|1} \right\rangle = \sum_{n=0}^3 \sum_{p=-\infty}^{\infty} \left\langle \tilde{A}^i_{n,p-n} \tilde{A}^j_{3-n,n-p} e_{n,i}^{0|1} , e_{3-n,j}^{0|1} \right\rangle = $$ $$ = \sum_{n=0}^3 \left\langle \mathcal{B}_n^{ij} e_{n,i}^{0|1} , e_{3-n,j}^{0|1} \right\rangle \ .
\end{equation}
It is now clear that the two BV symplectic forms could be equivalent after the identification
\begin{equation}\label{SEI}
	\sum_{n=0}^3 B_n^{ij} \equiv \sum_{n=0}^3 \mathcal{B}_n^{ij} \ \ , \ \ \text{i.e.} $$ $$ \sum_{n=0}^3 2 \left( \tilde{c}^i_{n,-n} \tilde{c}^{*j}_{3-n,n}  + \tilde{A}^i_{n,1-n} \tilde{A}^{*j}_{3-n,n-1} \right) \equiv \sum_{n=0}^3 \sum_{p=-\infty}^\infty \tilde{A}^i_{n,p-n} \tilde{A}^j_{3-n,n-p} \ .
\end{equation}
However, this equivalence would imply non trivial constraints on the fields, in particular it implies
\begin{equation}\label{SEJ}
	\sum_{\substack{p=-\infty \\ p \neq 0,1,2,3}}^\infty \tilde{A}^i_{n,p-n} \tilde{A}^j_{3-n,n-p} = 0 \ ,
\end{equation}
which is actually a non-trivial condition. Therefore, only under special conditions the field 
redefinition can be achieved and in general, this is not possible. 

Finally, the complete BV action for SCS at picture number 1, is given by the CS action with 
the complete sequence of fields

\begin{equation}\label{P1GFH}
	S_{SCS-BV} = \int_{\mathcal{SM}^{(3|2)}} Tr \left[ \frac{1}{2} \mathcal{A} \wedge d \mathcal{A} + \mathcal{A} \sum_{i=2}^{\infty} \frac{1}{i+1} \MM_i \left( \mathcal{A} , \mathcal{A} , \ldots \right) \right] \ , ~~~~ \mathcal{A} = 
	\bigoplus_{p = - \infty}^\infty A^{(p|1)}  \ .
\end{equation}

We may ask whether it is possible to select and construct a simplified action in terms of a sub-sector of fields which identifies a consistent, gauge invariant action. There are many examples in the literature (see e.g. \cite{Jurco:2018sby} and references therein) where it is shown that it is possible to build gauge invariant actions when working with $A_n$-structures or $L_n$-structures, i.e. algebraic structures including multiproducts up to $n$ entries (see \cite{Stasheff} for rigorous definitions). However, the multiproducts we are using do not come from an external structure (for example promoting the gauge Lie algebra to a non-associative version 
\cite{Gaberdiel:1997ia}) 
that we can freely modify, but they are defined in terms of the supermanifold geometry itself; therefore, we are not allowed to truncate the multiproducts with more than $n$ entries. Alternatively, we can investigate whether a sub-sector of our fields might lead to a gauge invariant action. There is one consistent choice, namely restricting ourselves to  the fields highlighted in \eqref{BCGC} only. They are proportional to $dx$ and the terms in the action involving higher products vanish; for example, we have
\begin{equation}\label{SEK}
	A_1^0 \wedge \MM_3 \left( A_1^0 , A_1^0 , A_1^0 \right) = \left( dx \right)^4 A_0^0 \wedge \MM_3 \left( A_0^0 , A_0^0 , A_0^0 \right) = 0 \ .
\end{equation}
That leads to a standard  BV Chern-Simons.  In that case, the condition (\ref{SEJ}) is satisfied by setting all coefficient 
to zero. Nonetheless, that solution can not be reached by a non-trivial field redefinition as shown above. 

Notice that from the previous arguments it follows that the restriction to the fields in \eqref{BCGC} is the only (evident) consistent possible restriction, since otherwise we would immediately need all the tower of possible fields as discussed above. Finally we conclude that the most general BV action for super-Chern-Simons theory is the one described in \eqref{P1GFH}.

\subsection{The Supersymmetric Term.}

As largely discussed in \cite{Nishino-Gates,Cremonini:2019aao} and as recalled in section 2, when considering super Chern-Simons theory with picture number 0, we have to add to the action the term $\displaystyle \left( W^{(0|0)} \right)^2 V^3 \wedge \mathbb{Y}^{(0|2)} $. This is necessary in order to have a closed Lagrangian. Recall that the requirement of a closed Lagrangian is related to the possibility of changing the PCO by $d$-exact terms, without modifying the action:
\begin{equation}\label{STA}
	\int_{\mathcal{SM}^{(3|2)}} \mathcal{L}^{(3|0)} \wedge \delta_V \mathbb{Y}^{(0|2)} = \int_{\mathcal{SM}^{(3|2)}} \mathcal{L}^{(3|0)} \wedge d \Lambda^{(-1|2)} = \int_{\mathcal{SM}^{(3|2)}} d \mathcal{L}^{(3|0)} \wedge \Lambda^{(-1|2)} = 0 \ .
\end{equation}
When considering the picture-1 case, the gauge field is not factorised any more, therefore we do not have explicitly the chance to choose the embedding, i.e. the PCO $\mathbb{Y}$. Moreover the Lagrangian is closed by definition, being a top integral form. Then the questions for the $W^2$ term raise: is this term necessary in the present case? How can we establish the connection with the usual picture-0 case? Again, the answer comes from the non-trivial structure of the picture-1 action. Let us focus on the kinetic term, in particular let us consider the $A_2^{-1} \wedge \left( d A \right)_1^1$. We can choose the fields such that
\begin{equation}\label{STB}
	A_2^{-1} = V^a \wedge V^b \epsilon_{abc} \iota \gamma^c W^{(0|1)} \ , \ \left( d A \right)_1^1 = V^a \psi \gamma_a W^{(0|1)} \ ,
\end{equation}
where $\left( d A \right)_1^1$ denotes the contribution to $F^{(1|1)}_1 = d A_2^{-1} $ 
when $d$ acts on $V^a$. Other terms are needed for covariance.  
These two terms together reproduce exactly the gaugino term of the super Chern-Simons action, with $W$ lifted to a form with picture number 1. This consideration highlights once again the generality and the non-trivial structure of the action written using the whole tower of forms.

\section{Conclusions}

We have studied BV super-Chern-Simons theory on supermanifolds at different picture numbers. The results 
are the following:
\begin{enumerate}
\item We built BV formalism for super-Chern-Simons starting from picture zero action. We pointed out that 
the natural space for antifields is given by the integral forms. There is a natural map between the components 
antifields and superspace antifields. That was never been noticed before and this gives additional strength to 
integral form formalism for any supersymmetric model or for any model built on supermanifolds. 
\item At picture number one, the model was described in \cite{Cremonini:2019aao} where it was enlightened the 
emergence of an $A_\infty$  structure as in super string field theory. Here, we show that the BV formalism is 
essential to deal with the intricacies of the gauge symmetry for multiproducts. 
\item We have shown that the richness of the multiproducts and non associative algebra is not avoidable. That has been 
proved by studying the BV symplectic form $\omega_{BV}$ and showing that there is no field redefinition which allows 
us to restrict the set of fields, and consequently the powers of multiproducts. In comparison with \cite{Gaberdiel:1997ia}, 
apparently here the $A_\infty$ algebra is intrinsically nested in the supermanifold geometry and cannot be 
redefined. That resembles how Haag-{\L}opusza\'nsky-Sonhius theorem overcomes the no-go theorem of Coleman-Mandula, 
since, as been noticed in  \cite{Gaberdiel:1997ia}, the non-trivial structures may emerge if they have a spacetime interpretation (see also \cite{Erler:2013xta}). 
\item We build the BV formalism for picture one gauge fields finding once more the beautiful geometric 
action of Chern-Simons for the entire set of fields and antifields as in string field theory. Furthermore, we 
show how the gaugino mass term is recovered from that action confirming the consistency of our results. 
The case of extended supersymmetry will be published elsewhere. 
\item Finally, it would be certainly very interesting to consider this theory a general supermanifold such 
as non-split/non-projected supermanifold \cite{Cacciatori:2017lky,Cacciatori:2017qyd,Noja:2018edj}.  
 
\end{enumerate}

\section*{Acknowledgements}
We are grateful to S. Cacciatori, R. Catenacci, L. Castellani, S. Noja, and C. Maccaferri 
for useful discussions during several stages of the research. The research is partially 
supported by ``Fondi di Ateneo per la Ricerca" by University of Eastern Piedmont. 

\appendix

\section{Notations}

We consider the case of a real supermanifold $\mathcal{SM}^{(3|2)}$; in terms of the coordinates, we define the following differential operators 
\begin{equation}
	\partial_a = \frac{\partial}{\partial x^a} \ , \ D_\alpha = \frac{\partial}{\partial \theta^\alpha} - \left( \gamma^a \theta \right)_\alpha \partial_a \ , \ Q_\alpha = \frac{\partial}{\partial \theta^\alpha} + \left( \gamma^a \theta \right)_\alpha \partial_a \ ,
\end{equation}
where the second and the third are known as superderivative and supersymmetry generator, respectively. They satisfy the superalgebra relations
\begin{eqnarray}
&&\left[ \partial_a , \partial_b \right] = 0 \, , \ \
\{D_{\a},D_{\b} \}=-2 \gamma^a_{\a\b} \partial_a\,, ~~~~
\{Q_{\a},Q_{\b} \}= 2 \gamma^a_{\a\b} \partial_a\,, ~~~~
\nonumber \\
&&
\{D_{\a},Q_{\b} \}=0\,,~~~~
\left\lbrace \partial_a , D_\alpha \right\rbrace = 0\,, ~~~~
\left\lbrace \partial_a , Q_\alpha \right\rbrace =0\ ,.
\label{susy3dB}
\end{eqnarray}
In 3d, for the local subspace we use the Lorentzian metric $\eta_{ab} = (-,+,+)$, and the real and symmetric Dirac matrices $\gamma^a_{\a\b}$ given by
\begin{eqnarray}
\label{dicA}
&&\gamma^0_{\a\b} = (C \Gamma^0) = - {\mathbf 1}\,, ~~~
\gamma^1_{\a\b} = (C \Gamma^1) = \sigma^3\,,  \nonumber \\
&&\gamma^2_{\a\b} = (C \Gamma^2) = - \sigma^1\,, ~~
C_{\a\b} = i \sigma^2 = \epsilon_{\a\b}\,. 
\end{eqnarray}
Numerically, we have $\hat\gamma_a^{\a\b} = \gamma^a_{\a\b}$ and 
$\hat\gamma_a^{\a\b} = \eta_{ab} (C \gamma^b C)^{\a\b} = C^{\a\gamma} \gamma_{a, \gamma\delta} C^{\delta\beta}$. 
The conjugation matrix is
$\epsilon^{\a\b}$ and a bi-spinor is decomposed as follows $R_{\a\b} = R \epsilon_{\a\b}  + R_a \gamma^a_{\a\b}$ where
$R = - \frac12 \epsilon^{\a\b} R_{\a\b}$ and $R_a = {\rm Tr}(\gamma_a R)$ are
a scalar and a vector, respectively.
In addition, it is easy to show that
$\gamma^{ab}_{\a\b} \equiv \frac12 [\gamma^a, \gamma^b]_{\alpha \beta} = \epsilon^{abc} \gamma_{c \a\b}$.

The differential of a generic function $\phi$ is expanded on a basis of forms as follows
\begin{equation}
d \phi = dx^{a} \partial_{a} \phi + d\theta^{\alpha}%
\partial_{\alpha}\phi =
\end{equation}
\[
=\Big(dx^{a} + \theta\gamma^{a} d\theta\Big) \partial_{a} \phi
+ d\theta^{\alpha}D_{\alpha}\phi \equiv V^{a} \partial_{a}
\phi + \psi^{\alpha}D_{\alpha}\phi\,,
\]
where $V^a = dx^a + \theta \gamma^a d\theta$ and $\psi^\a = d \theta^\a$ which satisfy the Maurer-Cartan 
equations 
\begin{eqnarray}
\label{DVDPSI}
d V^a = \psi \gamma^a \psi\,, ~~~~~~ d \psi^\a = 0\,. 
\end{eqnarray}

Given now a generic form $\Phi$, we can compute the supersymmetry variation and translation as a Lie derivative $\mathcal{L}_{\epsilon}$ with $\epsilon=\epsilon^{\a}Q_{\a}+\epsilon^{a}\partial_{a}$ ($\epsilon^{a}$ are the infinitesimal parameters of the translations and $\epsilon^{\a}$ are the supersymmetry parameters) and by means of the Cartan formula we have
\begin{equation}
\delta_{\epsilon}\Phi=\mathcal{L}_{\epsilon}\Phi%
=\iota_{\epsilon}d\Phi + d \iota_{\epsilon} \Phi=\iota_{\epsilon}\Big(dx^{a}\partial_{a}%
\Phi+d\theta^{\a}\partial_{\a}\Phi\Big) + d \iota_{\epsilon} \Phi =
\end{equation}%
\[
=(\epsilon^{a}+\epsilon\gamma^{a}\theta)\partial_{a}\Phi
+\epsilon^{\a}\partial_{\a}\Phi + d \iota_{\epsilon} \Phi =\epsilon^{a}\partial_{a}%
\Phi+\epsilon^{\a}Q_{\a}\Phi + d \iota_{\epsilon} \Phi\,,%
\]
where the term $ d \iota_{\epsilon} \Phi$ is simply a gauge transformation. It follows easily that $\delta_{\epsilon}V^{a} = \delta_{\epsilon}\psi^{\alpha}=0$ and  $\delta_{\epsilon}d \Phi = d \delta_{\epsilon}\Phi$.

\end{document}